\documentclass[12pt]{iopart}
\usepackage{iopams}
\usepackage{setstack}
\usepackage[dvips]{graphicx}
\usepackage{epsfig}

\begin{document}

\title{Random walk on a population of random walkers}

\author{E. Agliari$^1$, R. Burioni$^{1,2}$, D. Cassi$^{1,2}$, and F.M. Neri$^1$}
\address{$^1$ Dipartimento di Fisica, Universit\`a degli Studi di
Parma, viale Usberti 7/A, 43100 Parma, Italy}
\address{$^2$ INFN, Gruppo
Collegato di Parma, viale Usberti 7/A, 43100 Parma, Italy}

\begin{abstract}

We consider a population of $N$ labeled random walkers moving on a
substrate, and an excitation jumping among the walkers upon
contact. The label $\mathcal{X}(t)$ of the walker carrying the
excitation at time $t$ can be viewed as a stochastic process,
where the transition probabilities are a stochastic process
themselves. Upon mapping onto two simpler processes, the
quantities characterizing $\mathcal{X}(t)$ can be calculated in
the limit of long times and low walkers density. The results are
compared with numerical simulations. Several different topologies
for the substrate underlying diffusion are considered.

\end{abstract}


\section{\label{sec:intro}Introduction}

A general stochastic process $\xi(t)$ can be viewed as the time
evolution of one (or more) random variable \cite{vanKampen}, the
particular dependence on $t$ of the transition probabilities between
the states giving rise to different models. Among the most widely
studied stochastic processes in physics are Markov processes, where
the transition probabilities at $t_1>t$ depend only on $\xi(t)$ and
$t$, and not on the previous history of the system. In the simplest
case the time parameter $t$ is discrete, and $\xi(t)$ is called a
Markov chain; the case of a Markov chain with transition
probabilities independent of $t$ is by far the most studied. If 
the transition probabilities in the time interval
$(t_0,\,t_0+t)$ do depend on $t$ (with a given distribution
function), but not on $t_0$, we have \textit{homogeneous} processes.
Depending on the particular functional dependence on $t$, we can
obtain Poisson processes, Wiener processes, and so on. Relaxing the
homogeneity property, we can obtain the inhomogeneous version of the
previous processes.

Much more general assumptions on the time-dependence of the
transition probabilities can be given, but the resulting models are
rarely explicitly solvable. In this paper we define and solve a
particular discrete-time stochastic process: its transition
probabilities are a stochastic process themselves.


The process we consider is a ``second-level'' random walk, or
random walk on random walkers. We consider $N$ labeled random
walkers, diffusing on a given substrate. Such random walkers can
define a dynamic meta-graph: each random walk is seen as a node of
the meta-graph and a link between two of them is drawn whenever
they are within a distance $R$ on the substrate. Then, we study
the
diffusion of a ``second-level'' random walk on such meta-graph.\\
Apart from its mathematical interest, this kind of system is also
able to model a diffusion-reaction process. In fact, each walker
diffusing on the substrate represents a particle (all particles
belonging to the same chemical species) that can be either in an
excited ($A^*$) or in an unexcited ($A$) state, the former
corresponding to the node carrying the second-level random walker.
%
%
%
When an excited particle meets an unexcited one, they immediately
react according to the scheme
\begin{equation} \label{eq:reaction}
A^* + A \rightarrow A + A^*.
\end{equation}
This reaction mechanism is known as {\it homogeneous energy
transfer} (ET) which takes place from an excited molecule [donor
(A*)] to another unexcited molecule [acceptor (A)], according to
the scheme (\ref{eq:reaction}). This process stems from Coulombic
(long-range \cite{forster}) and exchange (non-radiative,
short-range \cite{dexter}) interactions amongst the particles. If
we just focus on the energy transfer via exchange (under the
implicit assumption that the relaxation takes zero time), this
allows to restrict transfer interaction to nearest-neighbour
particles only.

If we define an abstract space whose points are the $N$ random
walkers, the excitation transfer corresponds to a stochastic process
$\mathcal{X}(t)$ on the points of this space; hence, to a
``second-level'' random walk. The transition probabilities of this
process depend on the relative positions of the random walkers,
hence they are a stochastic process themselves. It is possible to
show how the process $\mathcal{X}(t)$ can be mapped exactly onto
simpler processes, involving $N$ or $N-1$ simple random walkers on
the same lattice; the study of the excitation jumps is here mapped
on the study of the passage times of these walkers through the
origin. These simpler processes can be solved in the limit of large
times and low walkers densities.

The paper is organized as follows. In Sec.~\ref{sec:Model} we
describe the model; in Sec.~\ref{sec:Analytical} we provide two
mappings to simpler processes that allow us to obtain the asymptotic
behaviour of the quantities of $\mathcal{X}(t)$. In Sec.~\ref{sec:Numerical}
these results are compared with numerical simulations. Sec.~\ref{sec:Conclusions}
contains our conclusions and perspectives.

\section{\label{sec:Model}The model}
We consider $N$ regular random walkers, labeled with the numbers
from 1 to $N$, moving on a finite structure (henceforth, the
substrate). The position of the $i$th walker at time $t$ is
$x_i(t)$; at time 0 all the positions are random. At $t=0$ one of the
walkers, $i_0$, carries an excitation; we assume without loss of
generality that $i_0=1$.

The following usual quantities for random walks on lattices will
be useful. For a walker starting from $r$ at time 0, we define the
probability $P_0(r,\,t)$ of being at 0 at time $t$, and the
probability $F_0(r,\,t)$ of being at 0 for the first time at time
$t$. We also define their generating functions,
$\tilde{P}_0(r,\,\lambda)=\sum_{t=0}^{\infty}P_0(r,\,t) \lambda^t$
and $\tilde{F}_0(r,\,\lambda)=\sum_{t=1}^{\infty}F_0(r,\,t)
\lambda^t$.

We fix a collision radius $R\geq 0$: at time $t$ two walkers meet
(or collide) if their distance on the lattice is $\leq R$. In this
paper we consider $R=1$ but there are no substantial differences for
different $R$ (the choice $R=0$ is
here neglected to avoid parity effects, and used
for explanations only in Sec. \ref{sec:Analytical}). When the walker $i$ carrying
the excitation collides with another walker $j$, the excitation
jumps from $i$ to $j$. If it collides with more than one walker at
the same time (which we will call a multiple hit), the excitation
jumps on one of them chosen randomly.

The model just described defines a discrete-time stochastic process
$\mathcal{X}(t)$, where the state space of the system is composed by
the set of the random walkers. At time $t$ the system is in state
$i$ if the excitation is on walker $i$

Formally, the process is defined by the state space

\begin{itemize}
\item $\mathcal{X}(t)\in\mathcal{N},\mathcal{N}=\{1,2,\ldots,N\}$;
\end{itemize}
by the initial condition:
\begin{itemize}
\item $\mathcal{X}(0)=1$;
\end{itemize}
and the evolution rule:
\begin{itemize}
\item let $\mathcal{X}(t)=i$; consider the set $\mathcal{C}=\{j:\|x_i(t)-x_j(t)\|\leq R;\,j\neq i\}$
\footnote{Here, $\|x-y\|$ denotes the chemical distance between $x$ and $y$, 
both for Euclidean and  fractal lattices.}.\\
If $\mathcal{C}=\varnothing$, then $\mathcal{X}(t+1)=i$.\\ If
$\mathcal{C}\neq\varnothing$, then $\mathcal{X}(t+1)=j$, where $j$
is chosen randomly among the elements of $\mathcal{C}$ with equal
probability.
\end{itemize}

Here, the transition (or jump) probabilities, given by the evolution
rule, are a stochastic process. In particular, at time $t$ the
transition probability from state $i$ ($\mathcal{X}(t)=i$) to state
$j$ ($\mathcal{X}(t+1)=j$) is a function of the positions $x_i(t)$
and $x_j(t)$ of the two RWs, hence a function of two stochastic
processes.

Several quantities can be defined for $\mathcal{X}(t)$, much in the
same way as for regular random walks on a lattice. We define:
\begin{itemize}
\item $\mathcal{J}(t)$, the average number of jumps performed by
the system up to time $t$; the probability $\mathcal{J}(t,\,h)$ that
the number of jumps performed by the system is $h$ at time $t$,
$\mathcal{J}(t)=\sum_{k=1}^{N} k \mathcal{J}(t,k)$. \item
$\mathcal{S}(t)$, the average number of different states visited at
time $t$; the probability $\mathcal{S}(t,k)$ that $k$ different
states have been visited by the system at time $t$,
$\mathcal{S}(t)=\sum_{k=1}^{N} k \mathcal{S}(t,k)$. \item the Cover
Time $\tau$, defined as the average time required to visit all the
$N$ walkers (analogous to the lattice-covering time for random walks
\cite{nemirovsky}). We also define $\pi$ as the average number of
jumps required to visit all the states ($\pi\leq\tau$).
\end{itemize}

The substrates considered will be Euclidean (hypercubic) lattices
of linear size $L$ and volume $L^d$ (with $d=1,\,2,\,3$), endowed
with periodic boundary conditions.

We also will consider fractal substrates. It is well known
\cite{havlin, burioni} that fractals are described by at least two
different dimensional parameters. One is the {\it fractal} dimension
$d_F$, describing the large-scale dependence of the volume (or mass)
$V(r)$ of the structure on the distance $r$ from a point 0 chosen as
the origin: $V_0(r)\sim a_0 r^{d_F}$ (here and in the following
lines, $a_0$, $b_0$ and $c_0$ are constants depending on the point
0). The other is the {\it spectral}, or connectivity, dimension
$d_s$, describing the long-time behaviour of diffusive phenomena on
the fractal. For example, for $t\rightarrow\infty$ the probability
of return to the starting point for a RW on the fractal is
$P_0(t)\sim b_0 t^{-d_s/2}$, and the average number of different
sites visited by the RW is $S(t)\sim c_0 t^{\min (d_s/2,\,1)}$. For
Euclidean lattices, $d_s=d_F=d$. In a lattice (either Euclidean or
fractal) with $d_s\leq 2$ a random walker starting from a point 0 is
bound to return to 0 an infinite number of times with probability 1,
and the lattice is called {\it recurrent}. For $d_s > 2$, the walker
has a non-null probability to escape to infinity without returning
to 0, and the lattice is called {\it transient}.

\begin{figure}[t]\begin{center}
\includegraphics[width=0.5\textwidth]{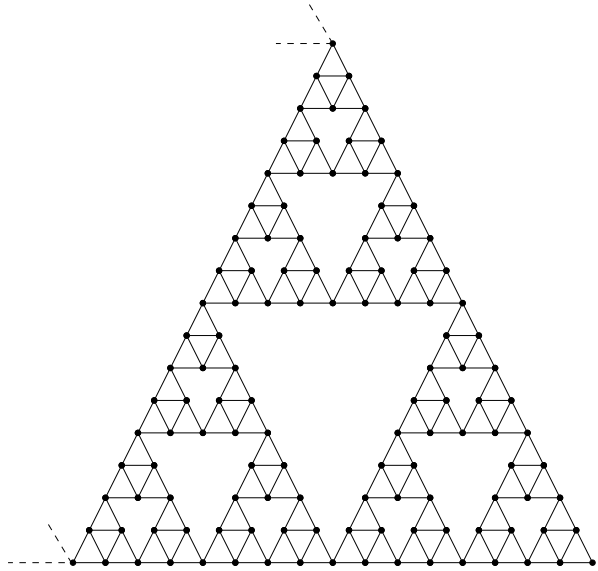}
\caption{\label{fig:gasket} Sierpinski gasket.} \end{center}
\end{figure}

The fractal lattices we will consider (fig. \ref{fig:gasket}) are
Sierpinski gaskets of linear size $L$ and volume $L^{\log 3/\log 2}$
($d_F=\log 3/\log 2$). Their spectral dimension is $d_s=2\log 3/\log
5$ (hence, they are recurrent: $d_s<2$) .

All the quantities we are interested in will be examined as
functions of $N$ and $L$.

\section{\label{sec:Analytical} Analytical Results}

The purpose of this section is to show how our model can be mapped
onto two different, and easier, models, that we shall call picture
1 and 2 respectively. In these two pictures, and in the
low-density (LD) limit (when multiple hits are negligible), the
asymptotic behaviour of the quantities of the previous section can
be found.

\begin{figure}[t]
\includegraphics[width=\textwidth]{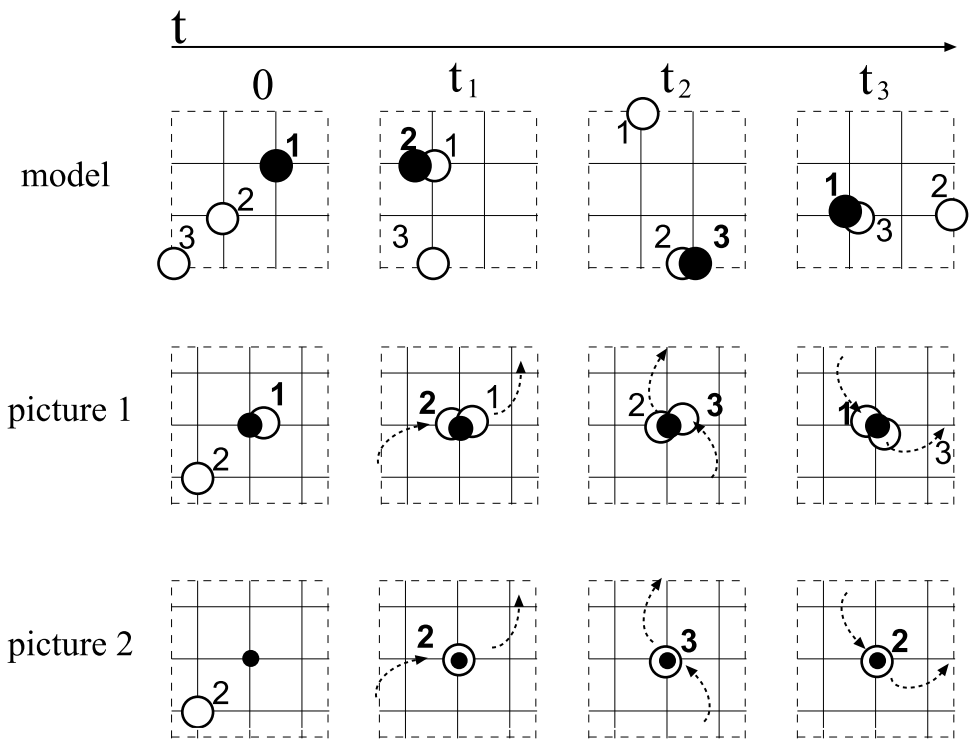}
\caption{\label{fig:mapping} {\it Top}:  the original process on a
square lattice at four nonconsecutive times 0, $t_1$, $t_2$, $t_3$.
The walker carrying the excitation is the black circle. The
excitation jumps from $1$ to $2$ at time $t_1$, from $2$ to $3$ at
time $t_2$, and from 3 to 1 at time $t_3$. {\it Middle}: picture 1.
The same process in the reference frame of the excitation (small
black circle fixed at the origin). The walkers are stuck at the
origin when carrying the excitation in the original model, and get
free when the excitation jumps to another walker. The jumping times
are the same. {\it Bottom}: picture 2. Here, the black circle marks
the origin. The \textit{associated free process}, with $N-1$ random
walkers labeled with the numbers from 2 to $N$, is shown. Picture 1
is obtained as follows. We start from the ordering $(1,\,2,\,3)$.
Each time walker $i$ of the associated process crosses the origin,
it exchanges its label with the previous walker that crossed it,
starting from walker 1; alternatively spelled, walkers at position 1
and $i$ of the present ordering exchange their labels. Hence, 2
crosses the origin at time $t_1$ and exchanges its label with 1; the
new ordering is $(2,\,1,\,3)$; 3 crosses the origin at time $t_2$
and exchanges the label with 2; the new ordering is $(3,\,1,\,2)$.
Finally, 2 crosses the origin at time $t_3$; the walkers at
positions 2 and 1 exchange their labels: the new ordering is
$(1,\,3,\,2)$.}
\end{figure}

Let us take figure \ref{fig:mapping} as a reference. The upper
part of the figure exemplifies the basic process. At $t=0$ the
excitation is on walker 1 (the system in state 1); at $t_1$ walker
2 hits walker 1 and the excitation jumps on walker 2 (the system
jumps on state 2). At times $t_2$ and $t_3$ the excitation jumps
on walker 3 and then on walker 1 again. This can be summarized by
introducing the sequence of jumping times
\begin{equation}\label{eq:seqTimes}
0,\,t_1,\,t_2,\,t_3,\,\ldots
\end{equation}
and the sequence of visited states
\begin{equation}\label{eq:seqStates}
0,\,i_1,\,i_2,\,i_3,\,\ldots
\end{equation}

{\bf Picture 1} ({\it stuck-and-free picture}) We consider the
process in the reference frame of the excitation. In this frame, the
walker carrying the excitation is stuck at the origin, and the other
$N-1$ walkers perform a regular random walk, with 2 jumps on each
time step. Here, the jump of the excitation from walker $i$ to
walker $j$ corresponds to the following: walker $j$ hits the origin
and gets stuck, while walker $i$ gets free and starts performing its
own RW.

In this picture, the process is a double-state RW process
\cite{weissLibro}, because each walker can exist in two different
states: either {\it stuck} at the origin or {\it free}. When a
walker is free, this picture allows us to use well-known quantities
from random-walk theory: for example, the probability for walker
$i$, starting from $r_i$ at time 0, of getting stuck at the origin
at time $t$ is (neglecting multiple hits) $F_0(r_i-r_0,\,2\,t)$.
This problem is still completely described by the above sequences of
times (\ref{eq:seqTimes}) and states (\ref{eq:seqStates}).


We remark that this mapping is possible only for translationally
invariant (Euclidean) lattices, where the lattice in the reference
frame of the excitation is the same as the original one. It is not
possible for fractal lattices; this will be clarified below.

{\bf Picture 2} ({\it label permutation picture})

When walker $2$ hits walker $1$ at the origin and gets stuck
(picture 1), the random walk subsequently performed by $1$ is just
the random walk that would have been performed by $2$ if no sticking
effect had existed: that is, if walkers $1$ and $2$ simply had
switched their labels without changing their state. This label
switch can be seen as the action of a transposition $(1\;2)$ of the
numbers 1 and 2 on the sequence $\mathcal{N}$.

Consider the process (let us call it the {\it associated free
process}) with $N-1$ free RWs, labeled from 2 to $N$, on the same
lattice, and walker 1 stuck once and for all at the origin. The
process in picture 1 is the same as the associated free process,
plus the following condition: when a walker hits the origin it
switches its label with the last walker that has hit the origin
before it (with the condition that the first walker has been 1).
In general, when walker $i$ of the associated free process hits
the origin, a permutation $\Pi = (1\;i)$ of elements $1$ and $i$
is induced on the original sequence $\mathcal{N}$ (since the last
stuck walker is always at the first place in the permutated
sequence).

The sequence of jump times (\ref{eq:seqTimes}) hence is equal to
the sequence of crossing times of $N-1$ random walkers through the
origin. Hence, the sequence of the walkers that cross the
origin in the associated free process
$$0,\,j_1,\,j_2,\,j_3,\,\ldots,$$
is related to the sequence for the original process by
$$i_1=(\Pi_1\mathcal{N})_{j_1};\,\,i_2=(\Pi_2\mathcal{N})_{j_2};\,\,i_3=(\Pi_3\mathcal{N})_{j_3};\,\ldots$$
where $\Pi_1=(1\;j_1)$; $\,\Pi_2=(1\;j_1)(1\;j_2)$; $\,\Pi_3=(1\;j_1)(1\;j_2)(1\;j_3)$, and so on.

Two observations are necessary at this point. First: both pictures
are valid only for translationally invariant lattices; for fractals,
for example, the lattice in the frame of reference of the excitation
does not coincide with the original one (indeed, it is not even
fixed but changes with $t$). However, several of our numerical
results suggest that the asymptotic results derived in the euclidean
case also hold (in some averaged sense) for non-integer-dimensional
cases. This point will be stressed again case by case.

Second: we depicted pictures 1 and 2 for a model with null range
$R=0$, while most of our numerical result concern the case $R\neq
0$ (mostly $R=1$), chosen to avoid the parity effects (since most of
our lattices are bipartite graphs, walkers starting from the
``wrong'' sites would never meet). A non-null range in the original
model corresponds to a sticking area greater than the origin in
picture 1, and to the passage to a region greater than the origin in
picture 2. This means that in picture 1 and 2 the walkers can perform jumps to the origin even when the origin is not a nearest-neighbor site. We expect, however, that the existence of a
non-null range will only result in a rescaling of the asymptotic
laws (usually by a factor $v/V$, where $v$ is the discrete volume of
the region). We will stress this point in the analytic results where
necessary.

\subsection{\label{sec:number of} Number of jumps for large times}

This quantity is easily calculated in picture 2. If we consider
low-density systems, that is, we neglect the probability of multiple
hits of the origin by the walkers, the number of jumps at time $t$
is the number of passages through the origin made by $N-1$ RWs at
time $t$, that is $N-1$ times the number of passages through the
origin made by a single RW. The mean number of times that a RW
starting from $r$ visits the origin in a walk of $t$ steps is
independent of $r$ for large $t$, and equals $\sim\frac{t}{V}$,
where $V$ is the volume of the lattice \cite{montroll}. The average
number of jumps is given by the mean number of times that $N-1$
independent RWs hit the origin, that is
\begin{equation}\label{eq:p_t}
\mathcal{J}(t)\sim\frac{N-1}{V}t,
\end{equation}
neglecting multiple hits. In the case of walkers with non-null
radius of action we must consider a finite-size trap. If $v$ is the
volume of the trap, the result is
\begin{equation}\label{eq:p_t2}
\mathcal{J}(t)\sim\frac{(N-1)\,v}{V}t.
\end{equation}
For example, for a radius $R=1$ we have $v=2d+1$ for hypercubic
lattices of dimension $d$.

For $\mathcal{J}(h,t)$ (the probability that the number of passages
performed by the excitation is $h$ at time $t$) no analytical
results are known, and we will rely only on numerical simulations.

\subsection{\label{sec:AverageFinal} Cover Time}

The Cover Time is defined as the average time needed for the system
to visit all the states. In the LD limit this is equal (looking at
picture 2) to the time needed for $N-1$ different walkers to be
absorbed into a trap located at the origin. This is a many-body
problem (already formulated in the frame of extreme value
statistics, see e.g. \cite{yuste}), and its exact solution is not
yet known.

We will adopt here an approximation. We recall that $F_0(r,\,t)$ is
the probability density for the first-passage time to the origin of
a walker starting from $r$. We know that on hypercubic lattices the
average first passage time for a RW through the origin, averaged
over all possible starting positions, is
$$\langle t \rangle_V =
\sum_r \frac{1}{V} \sum _{t=0}^{\infty} t F_0(r,t) \sim
a_d\,g_d(V),$$
where the approximation is valid for $V$ large; $a_d$ is a constant
that depends only on $d$, and $g_d(V)$ is the volume-depending part:
\begin{equation}\label{eq:trap_mon1}
g_d(V)= \left\{
\begin{array}{cc}
V^2 & d=1 \nonumber \\
V \, \log V & d=2 \nonumber \\
V & d>2
\end{array}\right.
\end{equation}
In the case of fractal lattices, the general formula $\langle t
\rangle_V\sim a_{d_s}\,g_{{d_s}}(V)$ can be heuristically justified,
and has been calculated analitically in two particular cases
\cite{kozak, agliari}; here,
\begin{equation}\label{eq:trap_mon2}
g_{{d_s}}(V)= \left\{
\begin{array}{cc}
V^{2/d_s} & d_s<2 \nonumber \\
V \, \log V & d_s=2 \nonumber \\
V & d_s>2,
\end{array}\right.
\end{equation}
$d_s$ being the spectral dimension of the lattice.

Our approximation consists in assuming that the first passage time
of the first out of $m$ RWs is that of one RW divided by $m$. Hence,
the time of absorption of the first walker is $g_d(V)/(N-1)$, that
of the second walker (the first out of $N-2$ left) is $g_d(V)/(N-2)$
and so on. The Cover Time is:
\begin{equation}\label{eq:tau_mon}
\tau(N,V) \sim \displaystyle \sum_{n=1}^{N-1}
\frac{a_d\,g_d(V)}{N-n} \sim \left[\gamma + \log N +
O(N^{-1})\right]\,a_d\,g_d(V),
\end{equation} where the
last relation holds in the limit of large $N$.

From what said before, we can easily estimate the average number of
jumps required to visit all the states:
\begin{equation}\label{eq:pi_mon}
\pi(N,V) = \displaystyle \frac{N-1}{V}\; \tau(N,V).
\end{equation}
In fact, as stated by equation (\ref{eq:p_t}), the average time
taken by the excited particle to meet another particle out of the
remaining $N-1$ is just $\frac{V}{N-1}.$

\subsection{\label{sec:Distinct}$\mathcal{S}(t)$, number of distinct particles visited at time $t$}

In the low-density limit (again looking at picture 2), this quantity
is the average number of particles (out of $N-1$) that survive at
time $t$ with a trap in the origin. This in turn is $N-1$ times the
survival probability of a single walker with a trap in the origin.

This quantity has been calculated in \cite{weiss} for Euclidean
lattices; let us quote here the main results. Let $U(t)$ and $S(t)$
be the survival probability of the walker and the average number of
sites visited by the walker at time $t$, respectively. The two
quantities are related by the formula $U(t)=1-S(t)/V$. Let
$S(\lambda)$ be the generating function of $S(t)$ with respect to
time. We have $S(\lambda)=f(\lambda)/(1-\lambda)$, where

$$f(\lambda)=\left[(1-\lambda)\phi(0,\lambda)+1/L^d\right]^{-1}.$$
The function $\phi(0,\lambda)$ constitutes the non-singular
contribution to the generating function $\tilde{P}_0(0,\,\lambda)$
as $\lambda \rightarrow 1$. More precisely, $\phi(0,\lambda)$ is
just a finite sum of terms involving the structure function of the
substrate.

The behavior of $f(\lambda)$ near its radius of convergence is
governed by $\phi(0,\bar{\lambda})$, where $\bar{\lambda}$ is the
root with the smallest magnitude of the equation
$f(\lambda)^{-1}=0$. For $d=1$ this value is known exactly to be
$\phi(0,\bar{\lambda})=2L/\pi^2$. For $d=2$, it is found numerically
that $\phi(0,\bar{\lambda})\sim 0.44 \log L$. For $d=3$,
$\bar{\lambda}=1$ and $\phi(0,\bar{\lambda})=1.51...$. Given these
results, the behavior of $U(t)$ for large times is

\begin{equation}\label{eq:DistinctStatic}
U(t) \sim \exp\left(-\frac{t}{L^d\phi(0,\bar{\lambda})}\right).
\end{equation}

We will find it expedient to write $U(t) \sim
e^{-\lambda_d\,t/g_d(V)}$ (cfr. equations (\ref{eq:trap_mon1}) and
(\ref{eq:trap_mon2})), where all the constants are absorbed in
$\lambda_d$. Hence,
\begin{equation}\label{eq:distinct}
\mathcal{S}(t)\sim (N-1)\left[ 1 -
\exp\left(-\frac{\lambda_d\,t}{g_d(V)}\right) \right].
\end{equation}
Now, by comparing $S(t) \sim V \left[ 1- U(t) \right]$ with
$\mathcal{S}(t)$ we can derive that the \textit{fraction} of
distinct particles excited $\frac{\mathcal{S}(t)}{N-1}$ just
corresponds to the \textit{fraction} $\frac{S(t)}{V}$ of distinct
sites visited by a regular random walker on the substrate. Equation
(\ref{eq:distinct}) holds also for fractals, replacing $d$ with
$d_s$.

For earlier times, the role of topology in the behavior emerges \cite{weiss}:
\begin{equation}\label{eq:early}
U(t) \sim \exp\left( -\frac{ \lambda_{d_s} \, t^{ min ({d_s}/2,1)
}} {g_{{d_s}}(V)} \right).
\end{equation}

Finally, notice that the (finite) size $R$ of the trap does not
qualitatively affect the previous relations while, in general, the
value of the constant $\lambda_{d_s}$ may non-trivially depend on
$R$. We will deepen this point later in Section
\ref{sec:distribution}.

\subsection{\label{sec:prob_distr} $\mathcal{S}(k,t)$, probability
distribution function for the $k$ distinct agents visited
at time $t$}

$\mathcal{S}(k,t)$ corresponds, in picture 2, to the probability
that the number of walkers absorbed into a trap at the origin is $k$.
Recalling that $U(t)$ is
the probability that a given walker has survived up to $t$, we have:
$$
\mathcal{S}(k,t)=U(t)^{N-k} (1-U(t))^{k-1}\left(
\begin{array}{cc}N-1\\k-1
\end{array}\right), \,\,\,\ 1 \leq k \leq N,
$$
that is (recalling that for Euclidean lattices $d_s=d$):
\begin{equation} \label{eq:prob_dist_trans}
\mathcal{S}(k,t)=e^{-(N-1) \lambda_{d_s} t/g_{{d_s}}(V)} \left(
e^{\lambda_{d_s} t/g_{{d_s}}(V)}-1 \right) ^{k-1}\left(
\begin{array}{cc}N-1\\k-1
\end{array}\right)
\end{equation}
Notice that, in the thermodynamic limit, equation
(\ref{eq:prob_dist_trans}) becomes a Poissonian distribution with
average $\mu = \lambda_{d_s} (N-1) t /g_{{d_s}}(V)$ (see
Fig.~\ref{fig:distribution}).

The time $t_{\mathrm{peak}}(k)$, each distribution is peaked at,
can be directly derived from equation (\ref{eq:prob_dist_trans}):
\begin{equation} \label{eq:t_peak}
t_{\mathrm{peak}}(k)=\frac{V}{\lambda_d} \log \left(
\frac{N-1}{N-k} \right).
\end{equation}

An important feature concerning
$\mathcal{S}(k,t_{\mathrm{peak}}(k))$ is that it exhibits a minimum
for $k=\tilde{k}=\frac{N+1}{2}$, as can be deduced from equations
(\ref{eq:prob_dist_trans}) and (\ref{eq:t_peak}).

It is as well possible to calculate the average time $\tau_{N-k}$
spent by the system having visited {\it exactly} $k$ different
states:
\begin{equation} \label{eq:t_k}
\tau_{N-k}= \displaystyle \sum_{t=0} ^{\infty}\mathcal{S}(k,t)
\sim \frac{V}{\lambda_{d_s}(N-k)},
\end{equation}
where the last relation was derived in the continuum limit for
$t$.

\begin{figure}[b!]
\includegraphics[width=.5\textwidth]{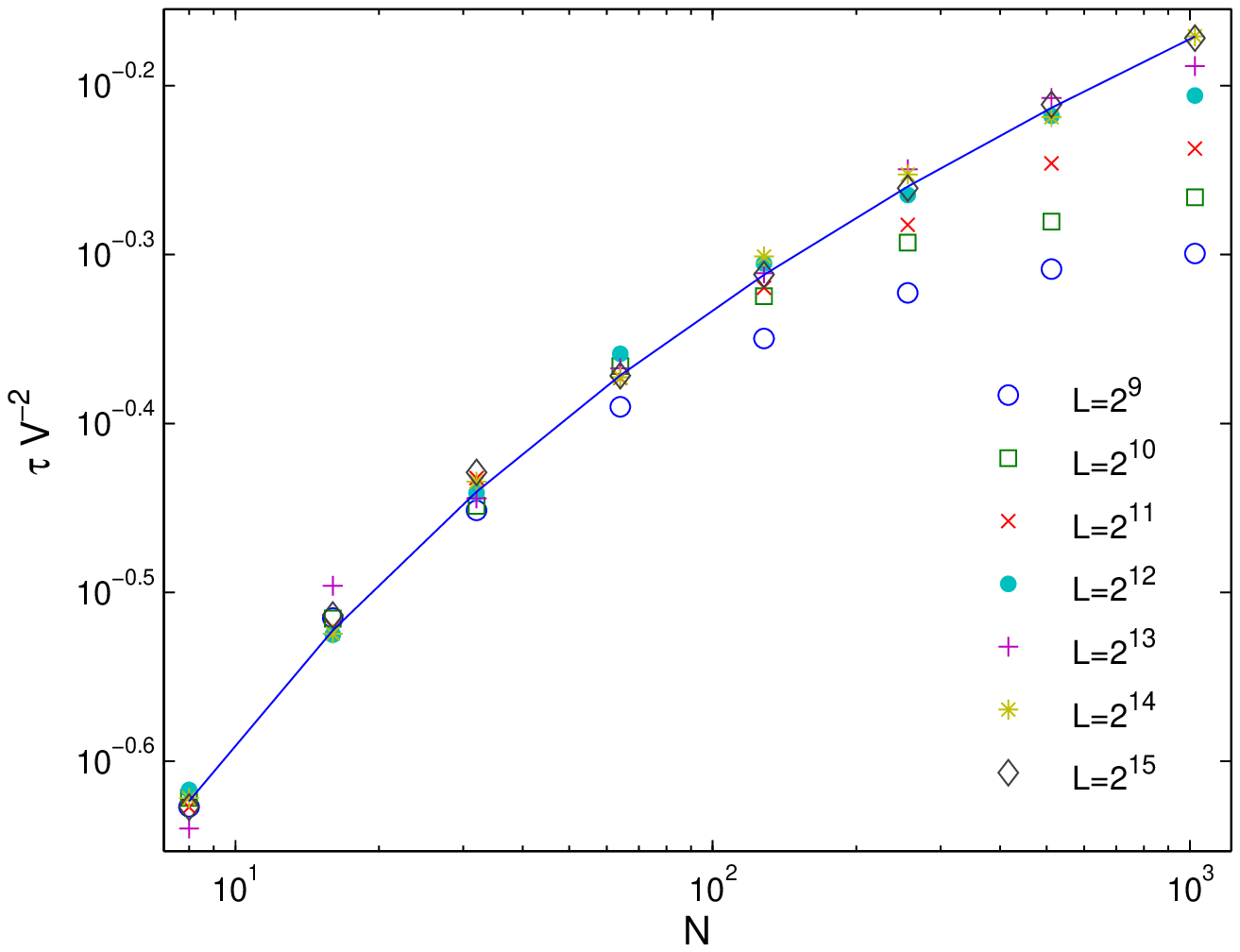}
\includegraphics[width=.5\textwidth]{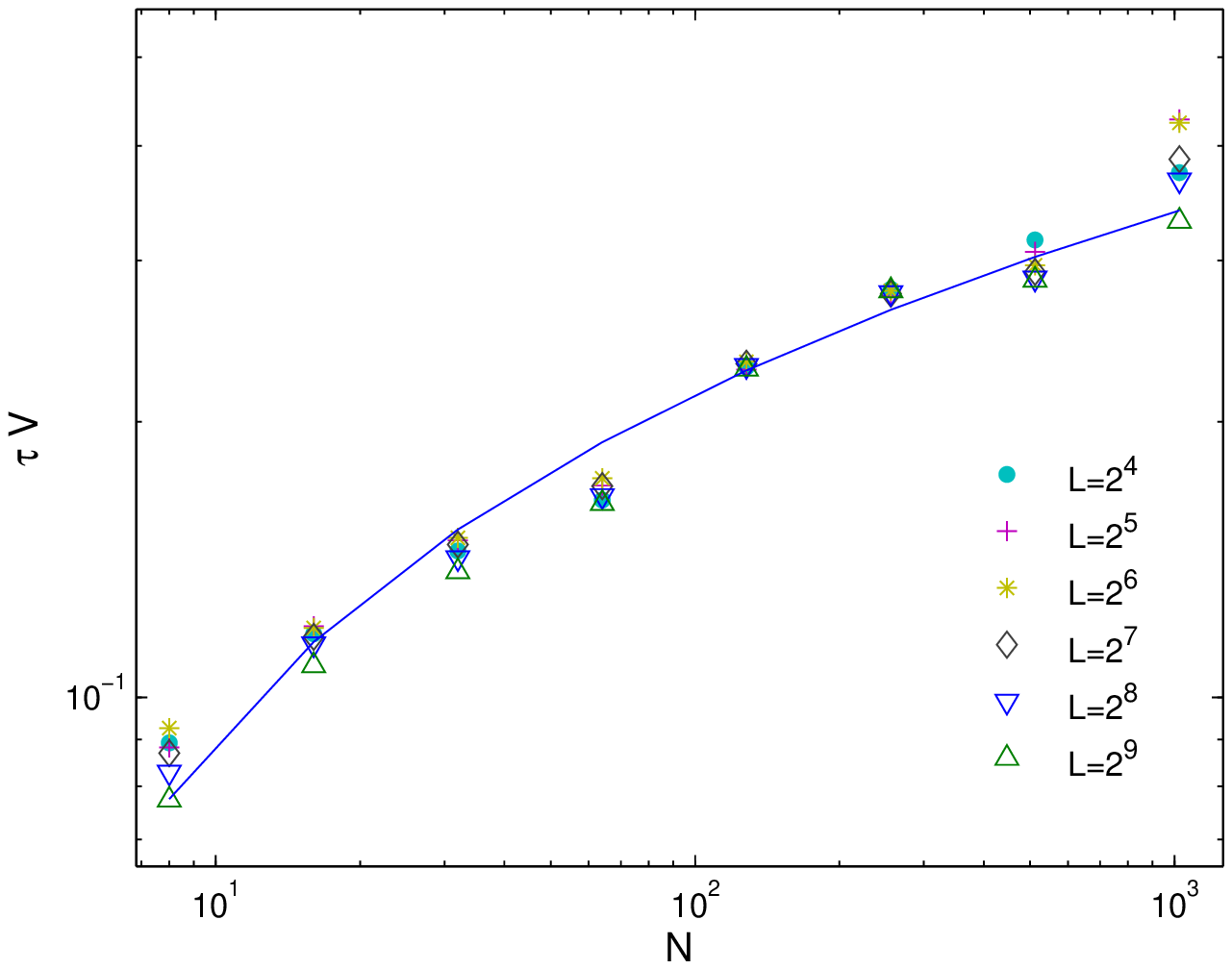}
\caption{\label{fig:Tau_mon} Rescaled Cover Time $\tau(N,V)$ versus
the number of walkers making up the system and diffusing on a
periodic chain (left panel) and cubic lattice (right panel).
Different sizes are considered, as shown by the legend. Equation
(\ref{eq:tau_mon}) provides the best fit when reactants
concentration is small.}
\end{figure}

\begin{figure}[b!]
\includegraphics[width=.5\textwidth]{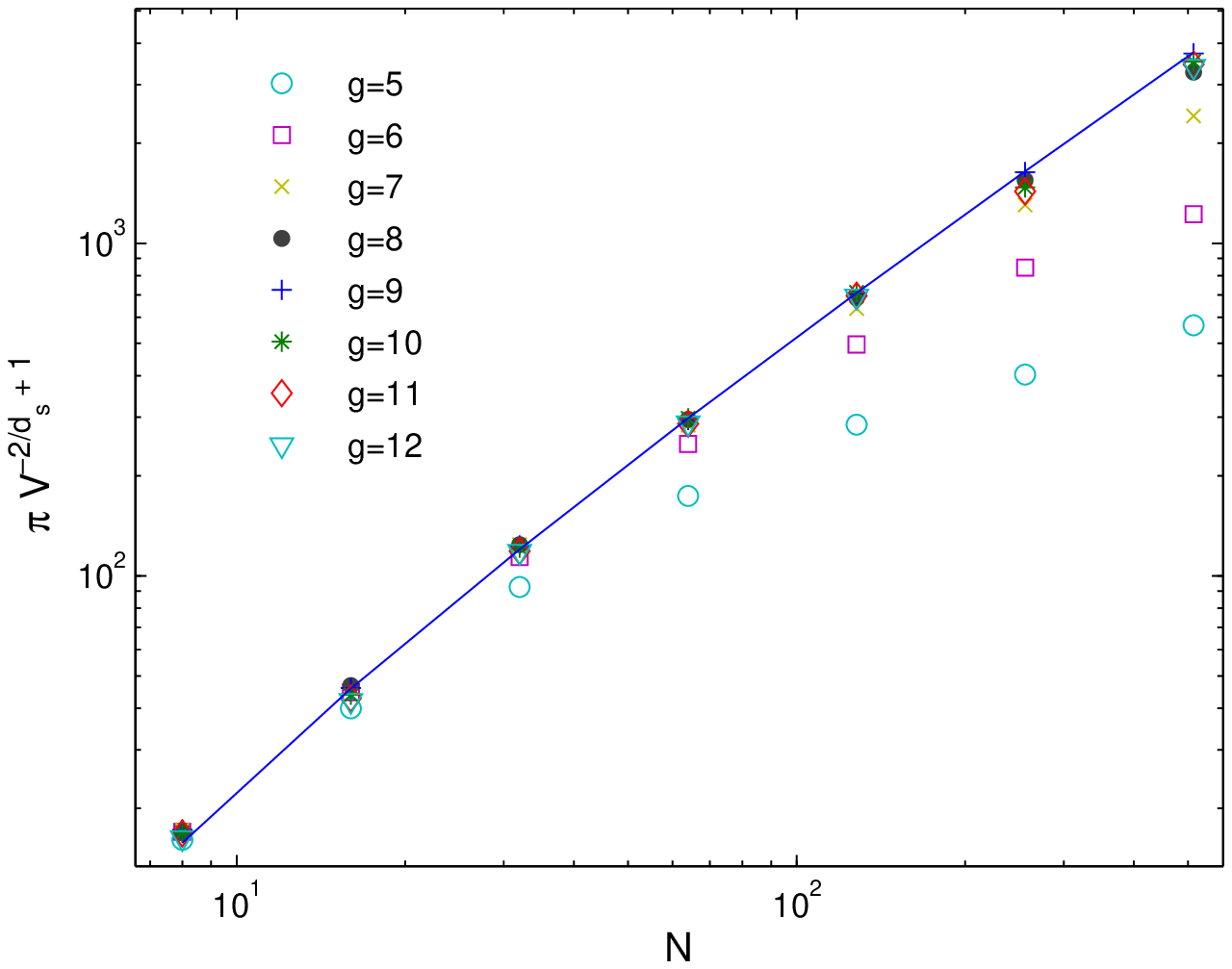}
\includegraphics[width=.5\textwidth]{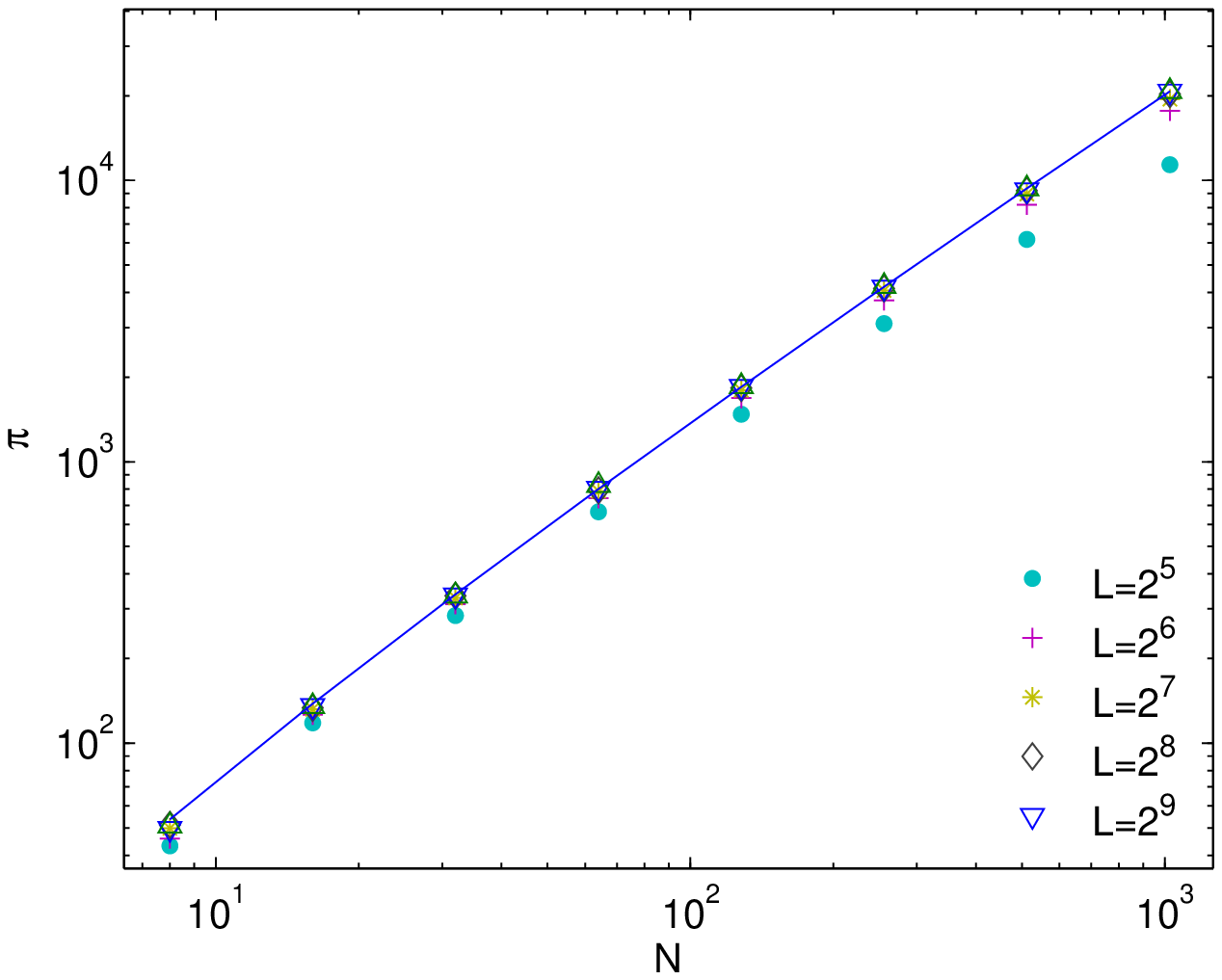}
\caption{\label{fig:Pi_mon} Rescaled Cover Jumps $\pi(N,V)$ for a
system of walkers diffusing and reacting on a Sierpinski gasket
(left panel) and on a cubic lattice with periodic boundary
conditions (right panel). Different sizes are depicted, as shown by
the legend. Equation (\ref{eq:pi_mon}) provides the best fit when the
reactants concentration is small. Notice that in the latter case
$\pi(N,V)$ is independent of $V$. }
\end{figure}

\section{\label{sec:Numerical}Numerical Results}

We first consider final quantities, i.e. quantities measured when
the excitation has covered the whole population of walkers. Subsequently,
we will take into account the temporal evolution of the system by
discussing quantities such as the average number of distinct
walkers $\mathcal{S}(t)$ visited at least once by the excitation,
as well as $\mathcal{S}(t,k)$ and $\mathcal{J}(t,k)$ representing
the probability distribution of having $k$ distinct walkers
visited at time $t$ and of having $h$ jumps performed at time $t$.

\subsection{\label{sec:final} Cover Time and Cover Jumps}

In this section we focus on numerical results concerning the Cover
Time $\tau$ and the Cover Jumps $\pi$. We recall that $\tau$ has
been defined as the average time it takes the excitation to reach
all the $N$ walkers diffusing on the substrate considered.
Analogously, $\pi$ represents the average number of jumps
performed by the excitation within the time at which
$\mathcal{S}=N$. Obviously, $\pi \leq \tau$.

In Figs.~\ref{fig:Tau_mon} and ~\ref{fig:Pi_mon} a proper rescaling
of data points confirms the analytical results discussed in the
previous section (see equations (\ref{eq:tau_mon}) and
(\ref{eq:pi_mon})). In particular, in the low-density regime,
$\tau(N,L)$ and $\pi(N,L)$ depend separately on $N$ and $L$ and
their functional form is strongly affected by the topology of the
lattice underlying the propagation (for example notice that for
transient substrates $\pi$ gets independent of the size of the
lattice).
%

%


\subsection{\label{sec:distribution} Distinct walkers Visited}
In Section \ref{sec:intro} we introduced $\mathcal{S}(t)$ as the
average number of distinct walkers which have been excited at
least once at time $t$.

\begin{figure}[b]\begin{center}
\includegraphics[width=.6\textwidth]{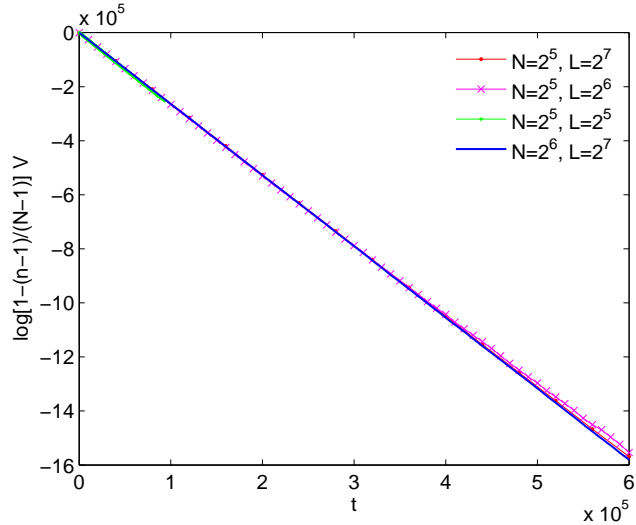}
\caption{\label{fig:S3} Rescaled number of distinct particles
visited by the second-level random walker as a function of time
for a periodic cubic substrate. Equation (\ref{eq:distinct}) holds
for any (low) concentration chosen. The only free parameter in the
fitting procedure is $\lambda_{d_s}^{fit}=2.65 \pm 0.05.$}
\end{center}
\end{figure}

In Sec.~\ref{sec:Analytical} we analytically showed that, in the
long-time regime, independently of the (finite) substrate topology
$\mathcal{S}(t)$ grows exponentially with time (see equation
(\ref{eq:distinct})). On the other hand, in the early-time regime,
for recurrent substrates, a functional dependence on the topology is
expected, consistently with what found for a random walker on a
finite lattice \cite{weiss}.

Let us first consider the case of a cubic structure for which the
behavior of $\mathcal{S}(t)$ is not expected to display any
crossover in time. Indeed, Fig.~\ref{fig:S3} confirms this: on the
whole range of time, equation (\ref{eq:distinct}) is a good
estimate for $\mathcal{S}(t)$ when the density is low. The slope
of $V \log \left(1- \frac{\mathcal{S}(t)}{N} \right)$ also allows
to derive an estimate for the constant $\lambda_{d_s}$. By fitting
numerical data we find that $\lambda_{3} \simeq 2.65(5)$,
$\lambda_{2} \simeq 6.84(4)$, $\lambda_{1} \simeq 10.01(8)$, (to
be compared with those in Sec.~\ref{sec:Distinct}, recalling that
here $R=1$).

\begin{figure}[b]\begin{center}
\includegraphics[width=.6\textwidth]{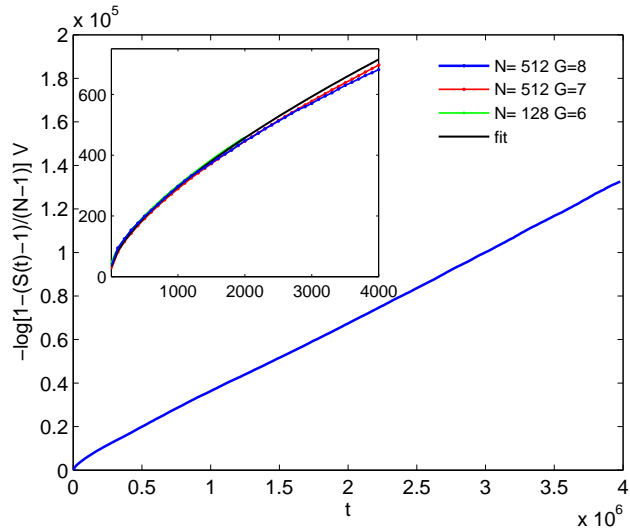}
\caption{\label{fig:SS} Time dependence for the number of distinct
walkers excited at least once and diffusing on Sierpinski gaskets
of different generations, as shown by the legend. The crossover
between the two time regimes is apparent by comparing the plot in
the large figure and the set of data depicted in the inset. The
quantity $-\log (1 - S(t)/N) V$ scales respectively as $t$ and
$t^{{d_s}/{2}}$.}
\end{center}
\end{figure}

Now, let us consider low-dimensional substrates.
The numerical simulations performed on the chain and on the
Sierpinski gasket (see Fig.~\ref{fig:SS}) support what previously
stated. In particular, for the latter we show that, at long time,
$\mathcal{S}(t)$ increases exponentially, analogously to what
previously found for the cubic lattice. Conversely, at small times,
deviations emerge: the pure-exponential growth is replaced by
$e^{t^{{d_s}/2}}$ in agreement with equation (\ref{eq:early}).

In Sec.~\ref{sec:intro} we introduced the function
$\mathcal{S}(k,t)$, representing the probability that, at time
$t$, the number of walkers visited at least once by the excitation
is $\mathcal{S}(t)=k$. In Sec.~\ref{sec:Analytical} we also
derived a mean-field approximation for this quantity, valid in the
low-density regime. We now discuss the pertaining results from
numerical simulations.

\begin{figure}[tb]
\begin{center}
\includegraphics[width=.6\textwidth]{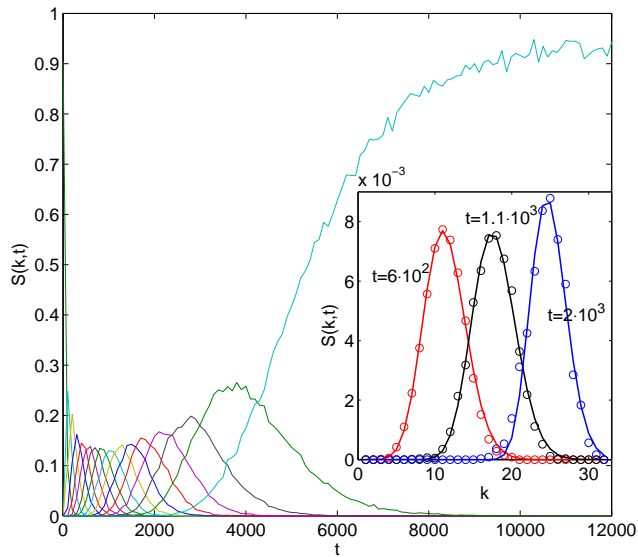}
\caption{\label{fig:distribution} Main Figure: Probability
distribution $\mathcal{S}(k,t)$ versus time $t$ for a system of
$N~=~32$ walkers diffusing on a cubic lattice sized $L=16$ with
periodic boundary conditions. Each curve represents a different
(even) value of $k$: starting from the leftmost distribution
$k=2,4,6,...,32$. Inset: Probability distribution
$\mathcal{S}(k,t)$ versus number of visited random walkers $k$;
three different instant of time are depicted in different colors:
$t=6 \cdot 10^2, 1.1 \cdot 10^3, 2 \cdot 10^3$. Data points
($\circ$) are fitted by a Poissonian distribution with average
$\mu_{{d_s}}=\lambda_{{d_s}} \rho t$ in agreement with what stated
in Sec.~\ref{sec:prob_distr}}.
\end{center}
\end{figure}

\begin{figure}[tb]
\includegraphics[width=.5\textwidth]{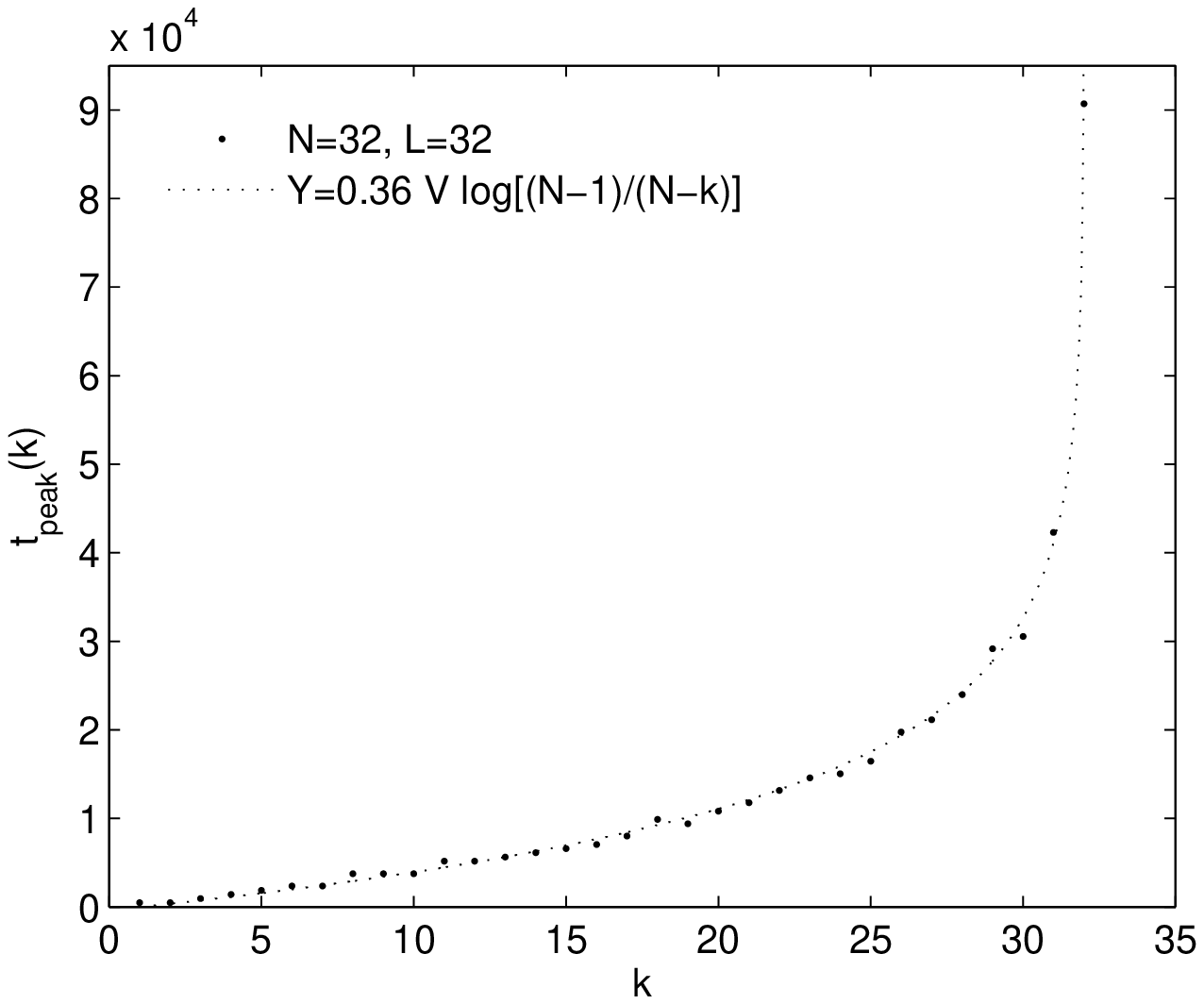}
\includegraphics[width=.5\textwidth]{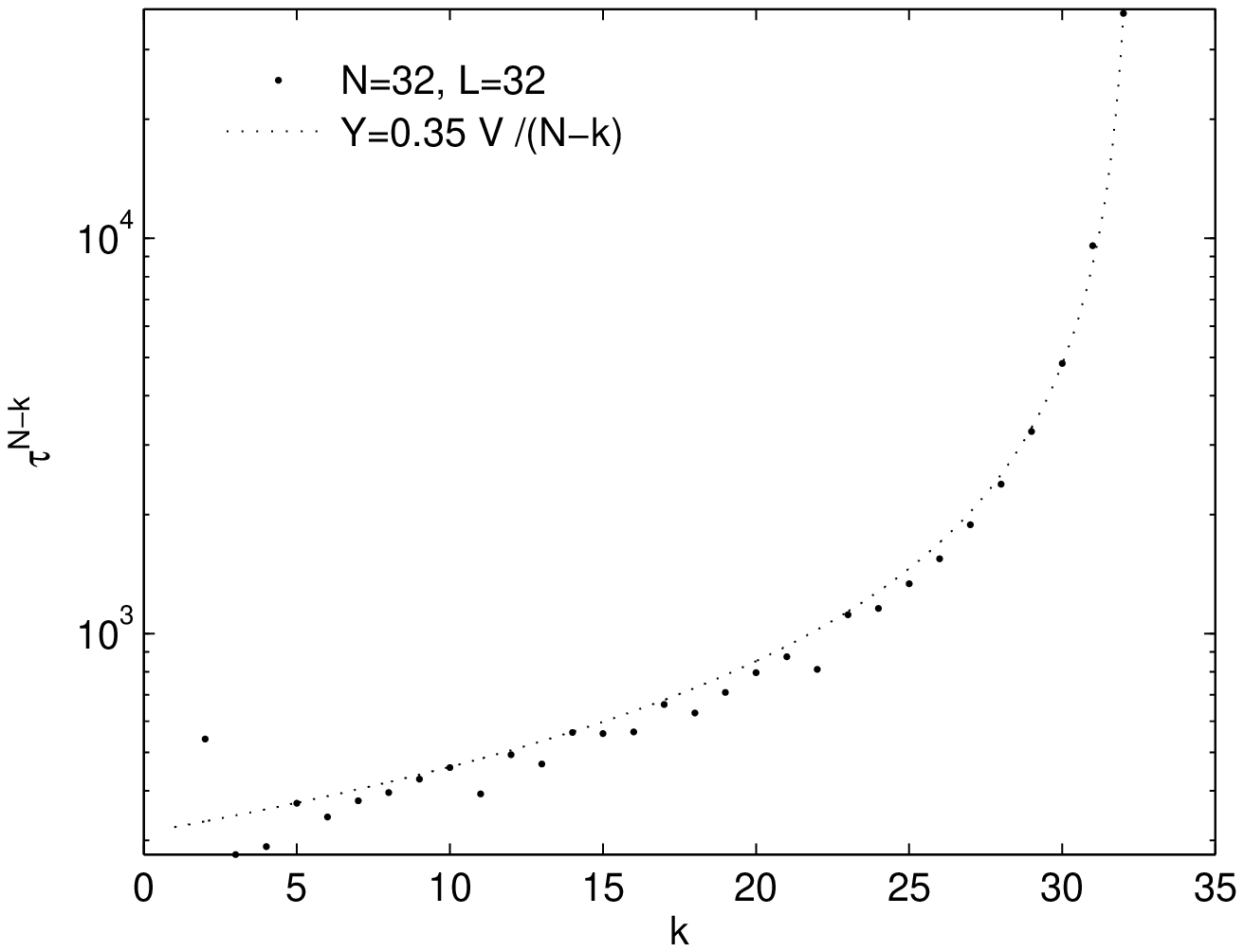}
\caption{\label{fig:tpeak} $t_{\mathrm{peak}}(k)$ and $\tau^{N-k}$
as a function of $k$ for a periodic cubic lattice. The dashed lines
(whose equations are reported) represent the best fits in agreement
with equations (\ref{eq:t_peak}) and (\ref{eq:t_k}). The only free
parameter is $\lambda_{d_s}$ and we get $\lambda_{d_s}^{fit}=2.79
\pm 0.07$.}
\end{figure}

In Fig.~\ref{fig:distribution} the probability distribution
$\mathcal{S}(k,t)$ is fitted by a Poissonian law with average $\mu$
linearly dependent on the density $\rho=\frac{N}{V}$ of the system.
Moreover, the time $t_{peak}$ each distribution is peaked at depends
on $k$ and it diverges logarithmically when $k \rightarrow N$ (see
Fig.~\ref{fig:tpeak}) according to equation (\ref{eq:t_peak}).

From the distribution $\mathcal{S}(k,t)$ it is also possible to
measure the average lifetime $\langle t_k \rangle$ for the $k$-th
state. This quantity diverges linearly as $k \rightarrow N$ as shown
in Fig.~\ref{fig:tpeak} where results for the cubic lattice are
depicted and fitted consistently with equation (\ref{eq:t_k}).

\begin{figure}\begin{center}
\includegraphics[width=.6\textwidth]{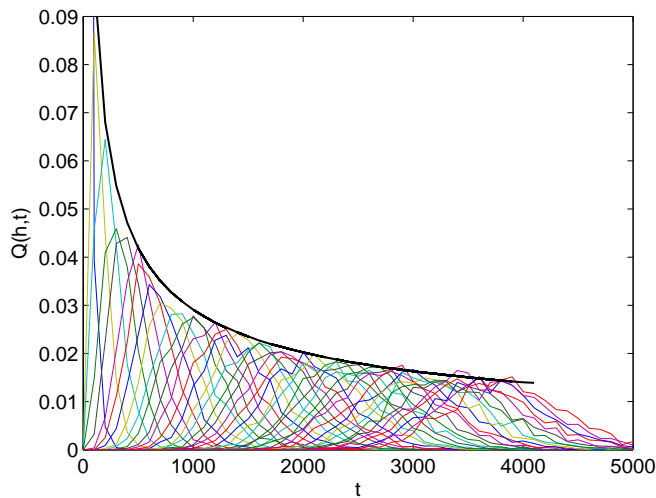}
\caption{\label{fig:distribution_pass} Probability distribution
$\mathcal{J}(h,t)$ versus time $t$ for a system of $N=32$ walkers
diffusing on a cubic lattice sized $L=16$, with periodic boundary
conditions. Several curves are depicted, each referring to a
different number of passages $h$ (selected one every 5 entries).
As $h$ increases, the extremal point of the related distribution
$t_{\mathrm{peak}}(h)$ gets larger, distributions are more and
more overlapped and fluctuations get more important. The best fit
for $\mathcal{J}(h,t_{\mathrm{peak}}(h))$ is represented by the
black line $y~=~A~t^B$, with $A=1.11 \pm 0.02, \, B=0.53 \pm
0.01$. Data have been averaged over $1.8 \cdot 10^5$
realizations.}
\end{center}
\end{figure}

An important feature emerging from Fig.~\ref{fig:distribution} is
the existence of a minimum for $\mathcal{S}(k,t_{peak})$. Indeed,
there exists a value $\tilde{k}$ at which the distribution is
maximally spread; in the average $\tilde{k}=\frac{N}{2}$ and,
correspondently, the statistical knowledge we have about the
system is minimum. From equation (\ref{eq:distinct}) we can
estimate $\tilde{t}\approx \frac{V}{\lambda_{d_s}} \log 2$.

Finally, in Fig.~\ref{fig:distribution_pass} numerical results for
$\mathcal{J}(h,t)$ are depicted. We recall that $\mathcal{J}(h,t)$
just represents the probability that the number
of passages performed by the excitation is $h$ at time $t$. From the
perspective of the energy-transfer mechanism this quantity is also
of practical interest, especially in the case we allow for energy
dissipation or emission during transfer. As shown in
Fig.~\ref{fig:distribution_pass}, there is no extremal point for
the envelop of such distributions which is indeed characteristic
of $\mathcal{S}(k,t)$.

\section{\label{sec:Conclusions} Conclusions and perspectives}
We have introduced and studied the diffusion of an excitation (or
second-level random walker) on a population of $N$ random walkers
diffusing on a given lattice (substrate) with finite volume $V$.
This results in a stochastic process $\mathcal{X}(t)$ whose
transition probabilities are themselves stochastic. The interest in
this kind of problem is also motivated by the fact that it provides
a model for systems of particles interacting by means of exchange
energy transfer.

We showed that in the low-density regime ($\rho=\frac{N}{V}\ll 1$)
$\mathcal{X}(t)$ can be mapped onto simpler processes, which allows
the analytic calculation of the quantities characterizing the
diffusion of the second-level RW. This analytic approach becomes
rigorous only for homogeneous substrates, but yields reliable
results also for fractal substrates. We presented numerical results
supporting our analytical findings.

There are two main possible developments for this model. First, one
can introduce a number $N_e >1$ of excitations jumping among the
walkers. This would allow for the existence of several donors
(excited walkers) in the system at the same time, and, possibly, of
several excitations residing on the same walker. The rules governing
the interaction between two donors (i.e., the existence of
constraints on the number of excitations on a single walker) would have to
be included in the model.

The second development consists in adding more levels of diffusion.
If we define a set of $N_e >1$ excitations, we obtain a set of $N_e
>1$ second-level stochastic processes. We can then define a
collision rule for those stochastic processes (for example, two of
them collide when the two excitations are on the same walker). Then,
we can introduce a third-level stochastic process by allowing a
third population of walkers diffuse on the second population (that
of the excitations). The interplay between the properties of the
second- and third-level stochastic processes (and a fourth-level
one, and so on) could then be studied.

\end{document}